\documentclass[12pt]{article}
\usepackage{texdraw}
\begin{document}

\begin{center}
{ \Large \bf Linear harmonic oscillator in spaces with degenerate metrics}
\end{center}

\begin{center}
N.A. Gromov \\
{\it Department of Mathematics, Komi Science Center,\\
 Ural Division, Russian Academy of Sciences, \\
Chernova str., 3a, 167982 Syktyvkar, Russia} \\
e-mail: gromov@dm.komisc.ru
\end{center}

\section*{Abstract}

With the help of contraction method we study the harmonic oscillator in spaces with degenerate metrics, namely, on Galilei plane and in the flat 3D Cayley-Klein spaces $R_3(j_2,j_3).$ It is shown that the  inner degrees of freedom are appeared which  physical dimensions are different from the dimension of the space.

{\it PACS:} 02.40.Dr; 45.05.+x; 45.20.Jj

{\it Keywords:} Harmonic oscillator, Degenerate metrics, Fiber space

\section{Introduction}

Recently a family of classical superintegrable systems was defined on the spaces of constant curvature
(or Cayley-Klein spaces) with nondegenerate metrics from classical  groups \cite{BHSS-03,HB-05}, 
as well as from quantum groups \cite{BHR-05-1,BHR-05-2}.
But among   Cayley-Klein spaces   there are  spaces with degenerate metrics, which are not considered in the these papers. Meanwhile, it is quite reasonable take one of the superintegrable systems, say harmonic oscillator, on the spherical space and study the chain of its contractions to  Euclid space and then to  Galilei space with degenerate metrics. Since contractions from the spherical space to the Euclid space for
different classical superintegrable systems are discussed in \cite{BHSS-03}--\cite{BHR-05-2}, in this paper we consider  contraction of harmonic oscillator from Euclid  to Galilei plane and its  behavior in the flat 3D Cayley-Klein spaces $R_3(j_2,j_3).$

\section{Geometrical properties of Galilei plane}

The metrics of three flat Cayley-Klein planes in Cartesian coordinates can be described in unified manner
\cite{G-90} as
\begin{equation}
ds^2=dx^2+j^2dy^2,
\label{1}	
\end{equation}
where parameter $j=1,\iota,i.$ Euclid plane with signature $(+,+)$ is obtained for $j=1,$ Minkowski plane with pseudoeuclidean metrics of signature $(+,-)$ is obtained for $j=i$ and nilpotent value of the parameter
$j=\iota, \; \iota^2=0, \; \iota/\iota =1$ correspond to Galilei plane with degenerate metrics of signature $(+,0).$
Galilei plane is the simplest fiber space with 1D base $\{x\}$ and 1D fiber $\{y\},$
therefore has two independent metrics: first in the base and second in the fiber
\begin{equation}
ds^2_b=dx^2,\quad  ds^2_f=\frac{1}{j^2}ds^2 |_{dx=0}=\frac{1}{j^2}\left(dx^2+j^2dy^2\right) |_{dx=0}=dy^2.	
\label{2}
\end{equation}

A bundle  of lines through a point on three Cayley-Klein planes has different properties relative to the plane automorphism \cite{P-65}.
On Euclid plane, any two lines of the bundle are transformed to each other by rotation  around the point. On Galilei plane, there is one isolated line  that  do not superposed with any other line of the bundle by Galilei boost. On Minkowski plane, there are two isolated lines that  are invariant with respect to Lorentz transformations.

\vspace{5mm}

\begin{figure}[h]
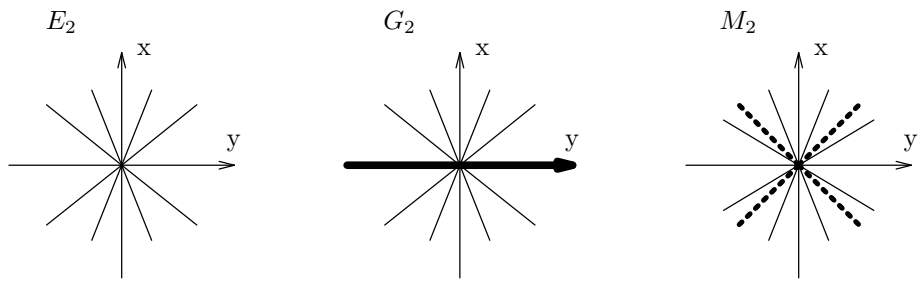

\centertexdraw{
\drawdim{mm}
\bsegment
\move(20 5)\arrowheadtype t:V \arrowheadsize l:2 w:1 \linewd 0.2 \ravec(0 30)
\move(5 20)\ravec(30 0)
 \move (20 20) \lvec(10 12)
 \move (20 20) \lvec(30 28)
 \move (20 20) \lvec(10 28)
 \move (20 20) \lvec(30 12)
 \move (20 20) \lvec(24 10)
 \move (20 20) \lvec(24 30)
 \move (20 20) \lvec(16 10)
 \move (20 20) \lvec(16 30)
\textref h:C v:C \htext(12 39) {\footnotesize $E_2$}
\textref h:C v:C \htext(23 35.5) {\footnotesize x}
\textref h:C v:C \htext(35 23) {\footnotesize y}
\esegment
\move(45 0)
\bsegment
\move(20 5)\arrowheadtype t:V \arrowheadsize l:2 w:1 \linewd 0.2 \ravec(0 30)
\linewd 1.0
\move(5 20)\ravec(30 0)
\linewd 0.2
 \move (20 20) \lvec(10 12)
 \move (20 20) \lvec(30 28)
 \move (20 20) \lvec(10 28)
 \move (20 20) \lvec(30 12)
 \move (20 20) \lvec(24 10)
 \move (20 20) \lvec(24 30)
 \move (20 20) \lvec(16 10)
 \move (20 20) \lvec(16 30)
\textref h:C v:C \htext(12 39) {\footnotesize $G_2$}
\textref h:C v:C \htext(23 35.5) {\footnotesize x}
\textref h:C v:C \htext(35 23) {\footnotesize y}
\esegment
\move(90 0)
\bsegment
\move(20 5)\arrowheadtype t:V \arrowheadsize l:2 w:1 \linewd 0.2 \ravec(0 30)
\move(5 20)\ravec(30 0)
\linewd 0.2
 \move (20 20) \lvec(10 14)
 \move (20 20) \lvec(30 26)
 \move (20 20) \lvec(10 26)
 \move (20 20) \lvec(30 14)
 \move (20 20) \lvec(24 10)
 \move (20 20) \lvec(24 30)
 \move (20 20) \lvec(16 10)
 \move (20 20) \lvec(16 30)
 \linewd 0.7
 \move (20 20)  \lpatt(0.5 1) \lvec(28 28)
 \move (20 20)  \lpatt(0.5 1) \lvec(12 12)
 \move (20 20)  \lpatt(0.5 1) \lvec(28 12)
 \move (20 20)  \lpatt(0.5 1) \lvec(12 28)
\linewd 0.2
\textref h:C v:C \htext(12 39) {\footnotesize $M_2$}
\textref h:C v:C \htext(23 35.5) {\footnotesize x}
\textref h:C v:C \htext(35 23) {\footnotesize y}
\esegment
}
\caption{A bundle  of lines on Euclid $E_2,$ Galilei $G_2$ and Minkowski $M_2$ planes.}
\end{figure}

  If one interpret these planes in some physical context, then on Euclid plane all lines must have the same
physical dimension $[x]=[y].$ On Galilei plane, there are infinite many lines with physical dimesion identical with  dimesion of the base $[x]$ and one isolated line in the fiber with some different physical dimesion $[y]\neq[x].$ On Minkowski plane, there are three types of lines, namely, $x$-like, $y$-like and zero-like ($x=\pm y$) that can be used for modelling of three physically different quantities.
The most known but not unique interpretations of Galilei and Minkowski planes are kinematical one, when $x$-axis is interpreted as time and $y$-axis is interpreted as space of (1+1) kinematics.

The flat 3D Cayley-Klein spaces $R_3(j_2,j_3)$ are defined by the metrics 
\begin{equation}
ds^2=dx^2+j_2^2dy^2+j_2^2j_3^2dz^2,
\label{3}	
\end{equation}
where parameters $j_2=1,\iota_2,i, \; j_3=1,\iota_3,i, \; \iota_2^2=\iota_3^3=0, \; \iota_2\iota_3=\iota_3\iota_2\neq 0, \; \iota_k/\iota_k =1, k=2,3$ \cite{G-90}. These spaces provide more possibilities for unification of different physical quantities within the bounds of one geometry. In particular, for
$j_2=\iota_2, \; j_3=\iota_3$ doubly fiber space $R_3(\iota_2,\iota_3)$ with two projections is obtained.
The first projection has 1D base $\{x\}$ and 2D fiber $\{y,z\},$ the second projection acts in 2D fiber $\{y,z\}$ and has 1D base $\{y\}$ and 1D fiber $\{z\}.$ There are three independent metrics
\begin{equation}
ds^2_b=dx^2,\quad  ds^2_{1}=\frac{1}{j_2^2}ds^2 |_{dx=0}=dy^2, \quad	
ds^2_{2}=\frac{1}{j_2^2j_3^2}ds^2 |_{dx=dy=0}=dz^2,
\label{4}
\end{equation}
 therefore three different physical quantities $[x]\neq[y]\neq[z]$
can be modelled by  the space $R_3(\iota_2,\iota_3).$

Unlike euclidean and  pseudoeuclidean geometries, where only one and three  physical quantities can be modelled, respectively,  fiber geometries enable to unify arbitrary many different physical quantities under appropriate dimensions and fibers.

\section{Harmonic oscillator in Galilei plane}

The  action for the linear harmonic oscillator in Euclid plane reads as
\begin{equation}
S^*=\int^{t_2^*}_{t_1^*}\left\{\frac{m^*}{2}\left(\dot{x}^{*2}+\dot{y}^{*2}\right)-\gamma^*\left(x^{*2}+y^{*2}\right) \right\}dt^*,
\label{5}
\end{equation}
where by star $^*$ are marked the initial euclidean variables.

To obtain the action for Galilei plane 
we use the method of unified description of Cayley-Klein spaces, groups, algebras etc. \cite{G-90}.  The main idea of this method is that construction suitable for all  Cayley-Klein cases can be obtained from an analogous construction for spherical space, orthogonal group, orthogonal algebra etc. by an appropriate transformation with the help of contraction parameters.

 Let us transform Cartesian coordinates as follows: $x^*=x, \; y^*=jy,$ then with 
 $t^*=t,\; m^*=m,\; \gamma^*=\gamma$  we obtain the action for the  harmonic oscillator  in the form
\begin{equation}
S=S^*(\rightarrow)=\int^{t_2}_{t_1}\left\{\frac{m}{2}\left(\dot{x}^{2}+j^2\dot{y}^{2}\right)-\gamma \left(x^{2}+j^2y^{2}\right) \right\}dt,
\label{6}
\end{equation}
where the arrow $(\rightarrow)$ means that transformed variables are substituted instead of initial ones.
For $j^2=\iota^2=0,$ that is for  Galilei plane, the one dimensional linear oscillator along the base $\{x\}$ is described by  the action 
\begin{equation}
S_b=\int^{t_2}_{t_1}\left\{\frac{m}{2}\dot{x}^{2}-\gamma x^{2} \right\}dt.
\label{7}
\end{equation}

In standard classical mechanics the time $t$ is the real continuous parameter associated with nondegenerate metrics. There are two independent metrics on Galilei plane, so it looks quite natural to introduce two
real continuous parameters. One of them associated with the metrics in the base as  the time $t$ and another one associated with the metrics in the fiber as  the "`time"' $\tilde{t}.$ To obtain the action for
the fiber let us transform in addition to Cartesian coordinates the "`time"' $t^*=j\tilde{t}$  and as before $m^*=m,\; \gamma^*=\gamma,$ then
\begin{equation}
S=\frac{1}{j}S^*(\rightarrow)=\int^{\tilde{t}_2}_{\tilde{t}_1}\left\{\frac{m}{2}\left(
\frac{1}{j^2}\left(\frac{dx}{d\tilde{t}}\right)^{2}+ \left(\frac{dy}{d\tilde{t}}\right)^{2} \right)-\gamma\left(x^{2}+j^2y^{2}\right) \right\}d\tilde{t}.
\label{8}
\end{equation}

As it is easy to see, the same action (\ref{8}) can be obtained by the mass renormalization  $m=j^2m^*$ with untouched "`time"' $t^*=\tilde{t},\; S=S^*(\rightarrow)$ and $L=L^*(\rightarrow),$ where $L$ is Lagrangian. Therefore this approach can be used when instead of an action a Lagrangian is regarderd. 
 For nilpotent value of the parameter $j=\iota$ this mass renorma\-li\-zation is similar to the mass renormalization in $\phi^4 $ quantum field theory (compare with Eq. (9.36) in  \cite{R-84})
\begin{equation}
m^2=m_1^2 \left (1+\frac{g}{16\pi^2\epsilon} \right )\stackrel{\epsilon \rightarrow 0}{\longrightarrow}
m^2=\frac{m_1^2}{\epsilon'},
\label{9}
\end{equation}
 where $m_1$ is physical mass, $m$ is unobserved mass and $\epsilon' \approx \iota^4. $ 

To avoid an indefinite expressions in the action (\ref{8}) it is necessary to put $dx=0, $ 
which define the fiber $x=x_0=const,$ then ${dx}/{d\tilde{t}}=0. $
 After that the action 
\begin{equation}
S_f=\int^{\tilde{t}_2}_{\tilde{t}_1}\left\{\frac{m}{2}
\left(\frac{dy}{d\tilde{t}}\right)^{2} -\gamma x_0^{2} \right\}d\tilde{t}
\label{10}
\end{equation}
describe the free "`motion"' $y=u_0 \tilde{t} + y_0$ in this fiber.  Here $u_0, y_0$ are integration constants.
So instead of one parameter harmonic oscillator  trajectory in Euclid plane we have two parameter family of   trajectories, that  fill up the band $2A\times R$ in Galilei $G_2$  plane  (see Fig.2).

\begin{figure}[h]
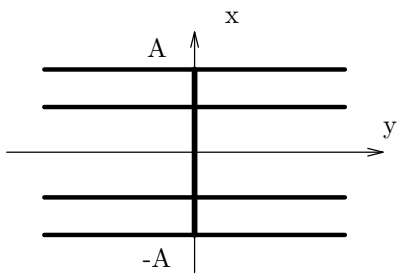

\centertexdraw{
\drawdim{mm}
\bsegment
\move(25 4)\arrowheadtype t:V \arrowheadsize l:2 w:1 \linewd 0.2 \ravec(0 32)
\move(0 20)\ravec(50 0)
\move(25 9) \linewd 0.8 \rlvec(0 22)
\textref h:C v:C \htext(30 38) {\footnotesize x}
\textref h:C v:C \htext(51 23) {\footnotesize y}
\textref h:C v:C \htext(20 34) {\footnotesize A}
\textref h:C v:C \htext(20 6) {\footnotesize -A}
\move(5 14)\linewd 0.6 \rlvec(40 0)
\move(5 26) \rlvec(40 0)
\move(5 9) \rlvec(40 0)
\move(5 31) \rlvec(40 0)
\linewd 0.2
\esegment
  } 
\caption{Two parameter family of harmonic oscillator trajectories in Galilei $G_2$  plane.}
\end{figure}

If one interpret the base as the space axis $[x]=[\mbox{space}],$   then  the fiber must have some different physical
dimension $[y]\neq[\mbox{space}]$ and can be regarded as some  inner degree of freedom.

\section{Harmonic oscillator in spaces $R_3(j_2,j_3)$}

The harmonic oscillator in three dimensional Euclid space $R_3$ is described by the action
\begin{equation}
S^*=\int^{t_2^*}_{t_1^*}\left\{\frac{m^*}{2}\left(\dot{x}^{*2}+\dot{y}^{*2}+\dot{z}^{*2}\right)-\gamma^*\left(x^{*2}+y^{*2}+z^{*2}\right) \right\}dt^*.
\label{11}
\end{equation}
The transition from $R_3$ to the space $R_3(j_2,j_3)$ is given by the following transformation of Cartesian coordinates
\begin{equation}
x^*=x, \quad y^*=j_2y, \quad z^*=j_2j_3z
\label{12}
\end{equation}
and the the action  is given by
\begin{equation}
S=S^*(\rightarrow)=\int^{t_2}_{t_1}\left\{\frac{m}{2}\left(\dot{x}^{2}+j_2^2\dot{y}^{2}+j_2^{2}j_3^2\dot{z}^{2}\right)-\gamma \left(x^{2}+j_2^2y^{2}+j_2^2j_3^2y^{2}\right) \right\}dt,
\label{13}
\end{equation}
where $t^*=t,\; m^*=m,\; \gamma^*=\gamma.$ 

Let $j_2=1,\; j_3=\iota_3,$ that is the space $R_3(1,\iota_3)$ has 2D base $\{x,y \}$ and 1D fiber $\{z \}.$
The two dimensional linear oscillator in the base  is described by  the action 
\begin{equation}
S_b=\int^{t_2}_{t_1}\left\{\frac{m}{2}\left(\dot{x}^{2}+\dot{y}^{2} \right)-\gamma \left( x^{2}+y^2 \right)\right\}dt,
\label{14}
\end{equation}
which is identical with (\ref{5})
and its trajectories are ellipses.
With $t^*=j_3\tilde{t} $ the action (\ref{11}) is transformed to
$$
S=\frac{1}{j_3}S^*(\rightarrow)=
$$
\begin{equation}
=\int^{\tilde{t}_2}_{\tilde{t}_1}\left\{\frac{m}{2}\left(
\frac{1}{j_3^2}\left( \left(\frac{dx}{d\tilde{t}}\right)^{2}+ \left(\frac{dy}{d\tilde{t}}\right)^{2}\right) + \left(\frac{dz}{d\tilde{t}}\right)^{2}\right)-\gamma\left(x^{2}+y^{2} +j_3^2z^{2}\right) \right\}d\tilde{t}.
\label{15}
\end{equation}
If we put $dx=dy=0, $ 
which define the fiber $x=x_0, y=y_0,$ then ${dx}/{d\tilde{t}}={dy}/{d\tilde{t}}=0 $ and the action 
\begin{equation}
S_f=\int^{\tilde{t}_2}_{\tilde{t}_1}\left\{\frac{m}{2}
\left(\frac{dz}{d\tilde{t}}\right)^{2} -\gamma \left(x_0^{2}+y_0^{2}\right) \right\}d\tilde{t}
\label{16}
\end{equation}
describe the free "`motion"' $z=w_0 \tilde{t} + z_0$ in this fiber.  
So we obtain two parameter family of   trajectories, that  fill up the elliptic cylinder  in  $R_3(1,\iota_3)$ (see Fig.3).

\begin{figure}[h]
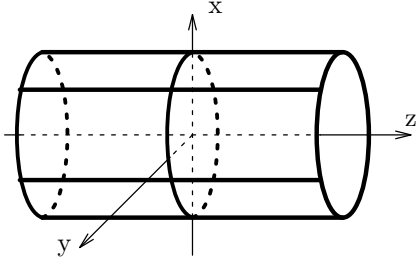

\centertexdraw{
\drawdim{mm}
\bsegment
\move(25 4) \linewd 0.2 \rlvec (0 5) \lpatt(0.5 1) \rlvec(0 22) \lpatt()
\arrowheadtype t:V \arrowheadsize l:2 w:1  \ravec(0 5)
\move(0 20)\rlvec(2 0)\lpatt(0.5 1) \rlvec(40 0) \lpatt() \ravec(12 0)
\move(25 20) \lpatt(0.5 1) \rlvec(-3 -3) \lpatt() \ravec(-12 -12)
\textref h:C v:C \htext(28 37) {\footnotesize x}
\textref h:C v:C \htext(8 5) {\footnotesize y}
\textref h:C v:C \htext(54 22) {\footnotesize z}
\move(2 14)\linewd 0.6 \rlvec(40 0)
\move(2 26) \rlvec(40 0)
\move(5 9) \rlvec(40 0)
\move(5 31) \rlvec(40 0)
\move(45 20)\lellip rx:3.5 ry:11
\lpatt()
\linewd 0.5
 \move(5 9) \clvec(0.5 10.5)(0.5 29.5)(5 31)
  \move(25 9) \clvec(20.5 10.5)(20.5 29.5)(25 31)
\lpatt(0.5 1.5)
 \move(5 9) \clvec(9.5 10.5)(9.5 29.5)(5 31)
 \move(25 9) \clvec(29.5 10.5)(29.5 29.5)(25 31)
\esegment
}
\caption{Two parameter family of harmonic oscillator trajectories in the space $R_3(1,\iota_3).$ }
\end{figure}

In the case of  doubly fiber space $R_3(\iota_2,\iota_3)$ according with (\ref{4}) the one dimensional linear oscillator along the main base $\{x\}$ is described by  the action (\ref{7}), the action for the base $\{y\}$
of the second projection is given by (\ref{10}) and the action for the last fiber $\{z\}$ is obtained with the help of time transformation $t^*=j_2j_3\hat{t} $ in the form
\begin{equation}
S=\frac{1}{j_2j_3}S^*(\rightarrow)|_{dx=dy=0}=\int^{\hat{t}_2}_{\hat{t}_1}\left\{\frac{m}{2}
\left(\frac{dz}{d\hat{t}}\right)^{2} -\gamma x_0^{2} \right\}d\hat{t}.
\label{17}
\end{equation}
As a result, we have three parameter family of   trajectories, that  fill up the region $2A\times R\times R$ in  the   
space $R_3(\iota_2,\iota_3).$
If one interpret the main base as the space axis $[x]=[\mbox{space}],$   then the base $\{y\}$ in the first fiber
and the second fiber $\{z\}$ all must have some different physical
dimension $[z]\neq[y]\neq[x]=[\mbox{space}]$ and can be regarded as two  inner degree of freedom of the system under consideration.

\section{Conclusion}

 We have considered the harmonic oscillator in fiber Cayley-Klein spaces which have several independent metrics in each base and in each fiber. Correspondingly, several actions and several real continuous parameters (or several "`times"')  are appeared. The action in the main base is obtained by the  coordinate transformations. 
The action in the fiber is obtained by the time transformation coupled with coordinate ones. In the case of Lagrangian the mass renormalization can be used instead of time transformation.
 The  fiber coordinates   can be interpreted as some inner degrees of freedom of the system which  physical dimensions are different from the dimension of the base coordinates.
In general, the  geometries with degenerate metrics are suitable  tools for unification of arbitrary many different physical quantities under appropriate dimensions and fibers.

\section*{Acknowledgements}
The author would like to thank J.F. Cari$\tilde{\mbox{n}}$ena, F.J. Herranz and V.V. Kuratov for helpful discussions.

\end{document}